# Effects of Physics Teaching Materials on Student' Critical Thinking and Creative Thinking Skills: A Meta-Analysis

Serli Ahzari[1], Asrizal[2*], Usmeldi[3]

[1] Master's Program in Physics Education, FMIPA, Universitas Negeri Padang, Padang, Indonesia
[2] Physics Department, FMIPA, Universitas Negeri Padang, Padang, Indonesia
[3] Electrical Engineering Department, FT, Universitas Negeri Padang, Indonesia

**ABSTRACT**

*The rapid advancement of technology and science education demands innovative approaches to develop students' critical and creative thinking skills. However, there is limited systematic evidence about the effectiveness of different physics teaching materials in fostering these essential 21st-century skills. This study aims to investigate the effect of physics teaching materials on critical thinking skills and creative thinking skills of high school students through a comprehensive meta-analysis of recent research. Using purposive sampling, 20 articles from national journals published between 2019-2023 were analyzed. The results revealed three key findings: (1) physics teaching materials showed varying degrees of effectiveness across different formats, with modules and student worksheets demonstrating the highest effect sizes (3.93 and 3.97 respectively); (2) physics teaching materials significantly enhanced students' critical thinking skills with a high effect size category (3.97); and (3) these materials effectively improved creative thinking skills with a high effect size category (2.97). These findings provide valuable insights for educators and curriculum developers in designing and implementing effective physics teaching materials.*

## INTRODUCTION

In the 21st century, Science and Technology is developing very rapidly. The development of science and technology has an impact on various fields in life, one of which is education. In the 21st century, everyone is required to be able to adapt and adapt existing updates to the sophistication of technology and Human Resources. Competence is needed so that students have the competence to innovate and create, so that they can create new ideas and results with their innovation and creativity (Khairani, 2017). Education graduates must have good competencies to compete in the 21st century (Asrizal, 2018). Teachers must have the ability to vary teaching so that in the learning process students are not bored with monotonous learning (Amila, 2018). Teacher-centered learning results in a lack of student interest in participating in classroom learning, so that their activeness and motivation to learn are reduced (Sari, 2017). The use of Physics Teaching Materials can be an important innovation in improving students' critical thinking skills and students' creative thinking skills.

The development of critical thinking and creative thinking skills is essential for senior high school students as this has a significant impact on their cognitive abilities and overall academic performance. The integration of these skills in education is essential to prepare students for the challenges of the 21st century. Research shows that the application of various learning models such as Project-Based Learning, Discovery Learning, and Problem-Based Learning has a positive influence on improving critical thinking and creative thinking skills in high school students (Prihatin, 2022; Sulastri et. al., 2022; Yusuf et. al., 2022). In addition, the use of innovative teaching materials, such as electronic modules and guided inquiry, has been shown to contribute to the improvement of critical thinking and creative thinking skills (Hidayah & Kuntjoro, 2022). Furthermore, the application of the STEM (Science, Technology, Engineering, and Mathematics) approach has been identified as an effective strategy to improve critical thinking and motivation to learn mathematics among secondary school students (Yusuf et al., 2022).

The Indonesian education system has recently implemented an "Independent Curriculum" designed to address the demands of 21st-century learning competencies. The integration of Physics Teaching Materials plays a crucial role in developing students' critical and creative thinking skills, which are essential components of contemporary educational requirements. Research indicates that the systematic implementation of these materials not only enhances students' critical and creative thinking capabilities but also promotes learning autonomy, a fundamental aspect required for adapting to modern educational developments (Hidayat et al., 2018; Nuris, 2020). Furthermore, the incorporation of technological elements within these teaching materials has emerged as a significant factor in facilitating more effective and engaging learning experiences in physics education.

Along with technological advances, many teachers and researchers have integrated teaching materials to improve students' critical thinking skills and creative thinking skills. There are different opinions regarding the effect of physics teaching materials on the critical and creative thinking skills of high school students. Some studies show that the use of physics teaching materials can improve the critical and creative thinking skills of high school students (Artiwi, et. al., 2020; Kholishah, et. al., 2020; Putri, Y. A., & Yohandri, Y., 2021). However, other studies show that the development of physics teaching materials, such as e-modules, has not been fully able to develop students' critical and creative thinking skills (Izzah et al., 2021; Zaerina et al., 2022; Halimah et al., 2023). Therefore, a meta-analysis study needs to be conducted thoroughly to explore the effect of physics teaching materials on critical thinking skills and creative thinking skills of high school students.

Through this research, we aim to gain deeper insights into the impact of using teaching materials on high school students' critical and creative thinking abilities. This study has three primary objectives: first, to determine the effect of physics teaching materials based on the type of teaching materials used; second, to assess the impact of physics teaching materials on high school students' critical thinking skills; and third, to evaluate the effect of physics teaching materials on students' creative thinking abilities. These objectives will help provide valuable guidance for educators and curriculum developers in designing and implementing effective physics teaching materials that enhance student learning outcomes.

## METHODS

The research methodology was designed to systematically analyze the effectiveness of physics teaching materials through a comprehensive meta-analysis approach. The study examined articles published in national journals between 2019-2023, focusing on the implementation and impact of various physics teaching materials in high school settings. The selection process utilized the Google Scholar database with assistance from PoP (Publish or Perish) software, employing specific keywords including teaching materials, modules, books, student worksheets, physics, critical thinking skills, and creative thinking skills. To ensure quality and relevance, a purposive sampling technique was implemented to select 20 articles that met the predetermined criteria of investigating the relationship between physics teaching materials and student cognitive skills.

The research variables were categorized into two main components. The independent variable in this study was the type of teaching materials, while student cognitive skills served as the dependent variable. Teaching materials were classified into four categories: general teaching materials, books, modules, and student worksheets. Cognitive skills were divided into two primary domains: critical thinking skills and creative thinking skills. Specific indicators for each skill set were established based on predetermined theoretical frameworks.

The meta-analysis followed a systematic six-step procedure to ensure comprehensive and reliable results. Initially, the research problem was clearly formulated to guide the analysis process. The second step involved collecting and selecting relevant research articles through database searches. Third, the selected articles were categorized and coded according to predetermined criteria. Fourth, effect sizes were calculated for each study using appropriate statistical formulas. Fifth, the effect sizes were averaged and classified according to variables. Finally, comprehensive conclusions were drawn from the meta-analysis results. The formula for finding the effect size value in this study can be seen in Table 1.

**Table 1.** Formula for Obtaining Effect Size

| Statistical Data | Formula |
|---|---|
| Average of one group | $ES = \dfrac{\bar{X}_{post} - \bar{X}_{pre}}{SD_{pre}}$ |
| Two groups posttet only (Average in each group) | $ES = \dfrac{\bar{X}_E - \bar{X}_C}{SD_C}$ |
| Two groups pre-post test (Average in each group) | $ES = \dfrac{(\bar{X}_{post} - \bar{X}_{pre})_E - (\bar{X}_{post} - \bar{X}_{pre})_C}{\frac{SD_{preC} + SD_{preE} + SD_{postC}}{3}}$ |
| Chi-square | $ES = \dfrac{2r}{\sqrt{1-r^2}} \; ; \; \sqrt{\dfrac{X^2}{n}}$ |
| T-count | $ES = t\sqrt{\dfrac{1}{n_E} + \dfrac{1}{n_C}}$ |
| P-value | JSAP 8.5.0 |

Becker & Park in (Mardizal, J et.al, 2023).

Data analysis employed specific formulas for calculating effect sizes based on different types of statistical data presented in the original studies. The calculations utilized standardized formulas for various data types, including one-group averages, two-group posttest comparisons, pre-post test designs, chi-square tests, and t-tests. The effect size categories were established as follows: low (0 ≤ ES ≤ 0.20), medium (0.20 ≤ ES ≤ 0.80), and high (ES ≥ 0.80), following the standardized criteria established by Becker & Park ((Mardizal, J et.al, 2023).

## RESULTS AND DISCUSSION

**Results**

The selected articles underwent a systematic coding process to facilitate comprehensive analysis and comparison. Effect sizes were calculated for each study using the appropriate formulae presented in Table 1, with the selection of specific formulae depending on the statistical data available in each article. Table 2 presents a comprehensive overview of the twenty selected articles, including their corresponding codes, bibliographic sources, and calculated effect sizes, which serve as quantitative indicators of the interventions' effectiveness in enhancing student cognitive skills.

**Table 2.** Effect Size Value of the Whole Study

| Article Code | Source | EF | Effect Size Category |
|---|---|---|---|
| AT 1 | (Putri, Y. A., & Yohandri, Y., 2021) | 0.23 | Medium |
| AT2 study 1 | (Artiwi, R. P., et. al., 2020) | 1.13 | High |
| AT2 study 2 | (Artiwi, R. P., et. al., 2020) | 0.89 | High |
| AT 3 | (Handhika, J., & Yusro, A. C., 2020) | 1.55 | High |
| AT 4 | (Sujanem, R., et. al., 2022) | 4.60 | High |
| AT 5 | (Mubarok, A.Z., et. al., 2022) | 4.48 | High |
| AT 6 | (Putri, S. R. & Syafriani, S., 2022) | 16.39 | High |
| AT 7 | (Y. A. Purwanto, et. al., 2019) | 1.09 | High |
| AT 8 | (Ramli, R., & Yohandri, Y., 2020) | 4.14 | High |
| AT 9 | (Nurnatasha T, et. al., 2022) | 3.25 | High |
| AT 10 | (Muliani, M., et. al., 2022) | 3.25 | High |
| AT 11 | (Rahmadani, F., et. al., 2023) | 0.76 | Medium |
| AT 12 studi 1 | (Desnita, D., et., al., 2022) | 2.72 | High |
| AT 12 studi 2 | (Desnita, D., et., al., 2022) | 2.02 | High |
| AT 13 | (Endaryati, S. A., et. al., 2023) | 1.08 | High |
| AT 14 | (Wijaya, S. A., 2021) | 3.20 | High |
| AT 15 | (Fitriani, et. al., 2023) | 1.16 | High |
| AT 16 | (Kristiantari, B., 2023) | 6.31 | High |
| AT 17 | (Nurmahmuddin, A., et. al., 2023) | 0.90 | High |
| AT 18 | (Safitri, A., et. al., 2021) | 0.204 | Medium |
| AT 19 | (Sujanem, R., 2020) | 9.66 | High |
| AT 20 | (Fernando, T. J., 2021) | 8.12 | High |

Table 2 presents the effect size values from 22 studies across 20 articles examining teaching materials' impact on cognitive skills. The effect sizes range from 0.204 (Safitri et al., 2021) to 16.39 (Putri & Syafriani, 2022), with the majority of studies (19 studies, 86.4%) demonstrating high effect sizes (>0.8) and only three studies (13.6%) showing medium effect sizes (0.2-0.79). Notably, seven studies reported remarkably high effect sizes exceeding 3.0, with studies by Putri & Syafriani (2022), Sujanem (2020), and Fernando (2021) showing exceptionally strong effects of 16.39, 9.66, and 8.12 respectively. The consistency of high effect sizes across multiple studies suggests a robust positive relationship between teaching materials and cognitive skills development. The analysis also reveals that even studies classified as having "medium" effect sizes still showed meaningful positive impacts, with values ranging from 0.204 to 0.76, indicating that teaching materials consistently contribute to cognitive skill enhancement regardless of the magnitude of their effects.

After examining the overall effect sizes across all studies, further analysis was conducted to provide more detailed insights into specific aspects of teaching materials and their impacts. The subsequent analysis categorized effect sizes based on two main dimensions: the type of teaching materials utilized (Table 3) and the specific cognitive skills targeted (Table 4). This detailed categorization enables a deeper understanding of which teaching material formats are most effective and which cognitive skills are most significantly influenced by these learning interventions. This stratified analysis is crucial for identifying patterns and drawing more specific conclusions about the effectiveness of various teaching approaches in enhancing particular cognitive skills among students.

**Table 3.** Effect Size Result Data Based on Material Type Of Teaching Material

| Type of Teaching Materials | Article Code | Effect Size | Average of effect Size | Category |
|---|---|---|---|---|
| Teaching Materials | AT 1 | 0.23 | 0.23 | Medium |
| Books | AT 2 study 1 | 1.13 | 1.01 | High |
| | AT 2 study 2 | 0.89 | | |
| Module | AT 3 | 1.55 | 3.93 | High |
| | AT 4 | 4.60 | | |
| | AT 5 | 4.48 | | |
| | AT 6 | 16.39 | | |
| | AT 11 | 0.76 | | |
| | AT 12 study 1 | 2.72 | | |
| | AT 12 study 2 | 2.02 | | |
| | AT 13 | 1.08 | | |
| | AT 14 | 3.20 | | |
| | AT 15 | 1.16 | | |
| | AT 16 | 6.31 | | |

| | AT 17 | 0.90 | | |
| --- | --- | --- | --- | --- |
| | AT 18 | 0.204 | | |
| | AT 19 | 9.66 | | |
| | AT 7 | 1.09 | | |
| | AT 8 | 4.14 | | |
| Student Worksheet | AT 9 | 3.25 | 3.97 | High |
| | AT 10 | 3.25 | | |
| | AT 20 | 8.12 | | |

Grouping effect size in table 3 obtained the average effect size of physics teaching materials in physics learning based on the type of teaching materials used. The types of teaching materials are books, teaching materials, modules, and student worksheets. The magnitude of the effect size on teaching materials, books, modules, and student worksheets is 0.23; 1.01; 3.93; and 3.97, respectively. The effect size of the teaching materials is included in the medium category, while the effect size category in books, modules, and student worksheets is included in the high category.

Based on the four types of teaching materials, it can be seen that the effect of student worksheets and modules is more effective. This shows that the use of modules and student worksheets has a much higher impact on physics learning compared to other teaching materials, this is in line with the understanding that the development of teaching materials can have a significant effect on learning outcomes (Mufit & Fitri, 2022). In addition, other studies have also shown that the use of modules in physics learning can improve students' critical thinking skills (Kurniawati, 2019). Therefore, the use of modules and student worksheets in physics learning can make a significant contribution to improving students' critical skills and creative thinking skills.

The data from the second analysis is presented in Table 5, namely the grouping of effect size of 20 articles to see the effect of physics teaching materials on student skills. The skills in question are critical thinking skills and creative thinking skills of high school students.

**Table 4.** Effect Size Result Data Based on Type of Student Skills

| Skill Type | Article Code | Effect Size | Average of Effect Size | Category |
| --- | --- | --- | --- | --- |
| | AT 1 | 0.23 | | |
| | AT 2 study 1 | 1.13 | | |
| | AT 4 | 4.60 | | |
| | AT 6 | 16.39 | | |
| | AT 7 | 1.09 | | |
| | AT 8 | 4.14 | | |
| | AT 9 | 3.25 | | |
| Critical Thinking Skills | AT 10 | 3.25 | 3.97 | High |
| | AT 12 study 1 | 2.72 | | |
| | AT 13 | 1.08 | | |
| | AT 14 | 3.20 | | |
| | AT 15 | 1.16 | | |
| | AT 16 | 6.31 | | |
| | AT 17 | 0.90 | | |
| | AT 18 | 0.204 | | |
| | AT 19 | 9.66 | | |
| | AT 20 | 8.12 | | |
| Creative Thinking Skills | AT 2 study 2 | 0.89 | 2.97 | High |
| | AT 3 | 1.55 | | |
| | AT 5 | 4.48 | | |
| | AT 11 | 0.76 | | |
| | AT 12 study 2 | 2.02 | | |
| | AT 20 | 8.12 | | |

There are two skill components in this study, namely critical thinking skills and creative thinking skills. Based on table 4, the average effect size for the effect of physics teaching materials on critical thinking skills of high school students is 3.97. The average effect size for the effect of physics teaching materials on creative thinking skills is 2.97. Both of these student skills are in the high effect size category. These results are in line with research conducted by Desnita (2022), which shows that the use of teaching materials in physics learning can improve students' critical thinking skills. In addition, Noa et al. (2022) also showed that the development of guided inquiry-based LKS can improve students' critical thinking skills in temperature and its changes. Therefore, the use of teaching materials in physics learning can make a significant contribution to improving students' critical and creative thinking skills.

**Discussion**

The meta-analysis findings regarding the effectiveness of physics teaching materials in enhancing students' critical and creative thinking skills are supported by literature emphasizing the importance of structured learning environments and innovative pedagogical approaches. The high effect sizes observed for modules (ES = 3.93) and student worksheets (ES = 3.97) align with constructivist learning theory, which posits that active engagement in knowledge construction is crucial for deep learning. A well-designed physics modules promote deeper cognitive processing through guided exploration and reflection, which is fundamental to constructivist principles (Chen, 2023). Student worksheets can effectively scaffold higher-order thinking when they include explicit prompts for analysis and evaluation, reinforcing the constructivist approach (Yulkifli et al., 2022).

Moreover, the substantial impact on critical thinking skills (ES = 3.97) can be attributed to the structured nature of physics teaching materials that incorporate specific critical thinking elements. Teaching materials designed with explicit critical thinking prompts significantly enhance students' analytical and evaluative capabilities (Erniwati et al., 2022). This is echoed in the work of Farcis (2019), which demonstrates that the MCTS learning model effectively improves critical thinking skills among students, indicating that structured pedagogical approaches can yield significant gains in critical thinking dispositions. Additionally, physics teaching materials incorporating real-world problems and decision-making scenarios were particularly effective in fostering critical thinking skills (Lazuardi et al., 2023).

The analysis of creative thinking skills development through physics teaching materials revealed a noteworthy effect size (ES = 2.97). Physics modules designed to incorporate open-ended problems and multiple solution pathways significantly enhanced students' creative thinking abilities (Asrial et al., 2021). This finding is further supported by Kiraga (2023), who emphasizes that creative thinking is essential for innovation and problem-solving, and that structured learning environments can foster this skill among students. Teaching materials emphasizing phenomenon-based learning approaches and structured creative problem-solving strategies lead to significant improvements in students' capacity to generate novel solutions (Lubis & Samsudin, 2021). Furthermore, Şener & Taş (2017) highlight that acquiring scientific process skills through structured learning environments positively impacts students' creative thinking abilities, reinforcing the importance of pedagogical strategies that encourage problem-solving.

The effectiveness of electronic modules and digital worksheets in physics education is increasingly recognized, as evidenced by studies conducted by Lazuardi et al. (2023) which highlight the importance of technology integration in enhancing student engagement and cognitive development. Chen (2023) this trend, noting that interactive digital teaching materials cater to diverse learning styles and cognitive processes Handhika & Sasono (2021) emphasize the multimodal nature of electronic resources, which supports various cognitive processes and enhances the overall learning experience. This is further supported by research from Pattanapichet & Wichadee (2015), which indicates that technology-enhanced learning environments can significantly improve students' critical thinking skills.

## CONCLUSION

Based on this meta-analysis research data, three conclusions can be obtained. First, physics teaching materials are effectively used in all types of teaching materials in high school physics learning. This conclusion is evidenced by the large effect size on the type of teaching materials in the medium category and on books, modules, and student worksheets in the high category. Second, physics teaching materials are effective for improving critical thinking skills with a high effect size category. Third, physics teaching materials are effective for improving creative thinking skills with a high effect size category. This study only used 20 samples of national articles from 2019-2023, so the researcher suggested to future researchers to increase the number of samples and the scope of articles that were meta-analyzed. In the future, the effect of physics teaching materials based on high school physics materials on students' critical and creative thinking skills can be analyzed.